\documentclass[conference]{IEEEtran}

\usepackage[T1]{fontenc}
\usepackage[utf8]{inputenc}
\usepackage{amsmath}
\usepackage{booktabs}
\usepackage{graphicx}
\usepackage{tabularx}
\usepackage{xcolor}
\usepackage{url}
\usepackage{microtype}
\usepackage{enumitem}
\usepackage{pdfpages}

\begin{document}

\title{Poster: ClawdGo: Endogenous Security Awareness Training\\for Autonomous AI Agents}

\author{
\IEEEauthorblockN{Jiaqi Li\IEEEauthorrefmark{1}\IEEEauthorrefmark{2},
Yang Zhao\IEEEauthorrefmark{1}\IEEEauthorrefmark{2},
Bin Sun\IEEEauthorrefmark{3},
Yang Yu\IEEEauthorrefmark{4},
Jian Chang\IEEEauthorrefmark{5},
Lidong Zhai\IEEEauthorrefmark{1}\IEEEauthorrefmark{2}}
\IEEEauthorblockA{\IEEEauthorrefmark{1}Institute of Information Engineering, Chinese Academy of Sciences, Beijing, China}
\IEEEauthorblockA{\IEEEauthorrefmark{2}School of Cyber Security, University of Chinese Academy of Sciences, Beijing, China}
\IEEEauthorblockA{\IEEEauthorrefmark{3}Network Management Center, China Mobile Group Liaoning Company Limited, Liaoning, China}
\IEEEauthorblockA{\IEEEauthorrefmark{4}Tencent Security Xuanwu Lab, Haidian District, Beijing, China}
\IEEEauthorblockA{\IEEEauthorrefmark{5}China Unicom Online Information Technology Co., Ltd., Beijing, China}
\IEEEauthorblockA{\texttt{zhailidong@iie.ac.cn}}
}

\maketitle

\begin{abstract}
Autonomous AI agents deployed on platforms such as OpenClaw face prompt
injection, memory poisoning, supply-chain attacks, and social engineering---yet
existing defences address only the platform perimeter, leaving the agent's own
threat judgement entirely untrained. We present \textbf{ClawdGo}, a framework
for \emph{endogenous} security awareness training: we teach the agent to
recognise and reason about threats from the inside, at inference time, with no
model modification. Four contributions are introduced: \textbf{TLDT}
(Three-Layer Domain Taxonomy) organises 12 trainable dimensions across
Self-Defence, Owner-Protection, and Enterprise-Security layers; \textbf{ASAT}
(Autonomous Security Awareness Training) is a self-play loop where the agent
alternates attacker, defender, and evaluator roles under weakest-first
curriculum scheduling; \textbf{CSMA} (Cross-Session Memory Accumulation)
compounds skill gains via a four-layer persistent memory architecture and Axiom
Crystallisation Promotion (ACP); \textbf{SACP} (Security Awareness Calibration
Problem) formalises the precision-recall tradeoff introduced by endogenous
training. Live experiments show weakest-first ASAT raises average TLDT score
from 80.9 to 96.9 ($\Delta{+}15.9$, 16 sessions), outperforming uniform-random
scheduling by 6.5 points and covering 11/12 dimensions. CSMA retains the full
gain across sessions; cold-start ablation recovers only 2.4 points
(13.6-point gap). E-mode generates 32 TLDT-conformant scenarios covering all
12 dimensions. SACP is observed when a heavily trained agent classifies a
legitimate capability assessment as prompt injection (30/160).
\end{abstract}


\section{Introduction}

The AI agent ecosystem has grown faster than its defences. OpenClaw, an
open-source autonomous agent framework released in November 2025, accumulated
over 250,000 GitHub stars within 60 days and has over 135,000 publicly
accessible instances on the internet~\cite{openclaw2026wiki,bitdefender2026}.
Snyk's 2026 ToxicSkills audit scanned 3,984 ClawHub and skills.sh packages and
found that 1,467 skills (36.82\%) had at least one security issue, including 76
confirmed malicious payloads validated through human review~\cite{snyk2026}.
SecurityScorecard reported over 40,000 internet-exposed OpenClaw instances, with
15.2K flagged as vulnerable to remote code execution in its updated exposure
analysis~\cite{securityscorecard2026}. NVD records CVE-2026-25253 as a CVSS 8.8
OpenClaw token-leakage vulnerability and CVE-2026-32922 as a CVSS 9.9
privilege-escalation-to-RCE vulnerability~\cite{cve25253,cve32922}.

Beyond platform vulnerabilities, social engineering campaigns impersonate
owners to redirect agent behaviour; memory poisoning corrupts trusted session
context; supply-chain attacks embed malicious behaviour in reputable skills.
What unites all these vectors is that they target \emph{the agent's own
judgement}---its trust in instructions, in memory, and in skill provenance.
No platform-level control addresses this attack surface.

Existing countermeasures harden the platform boundary: static scanners, runtime
filters, and sandboxing. These are necessary but insufficient. An unaugmented
agent confronting a CFO-impersonation payment request has no trained basis to
identify the authority-urgency-bypass pattern. A malicious skill requesting SSH
key access under the guise of a security patch exploits the agent's lack of
supply-chain threat intuition. Endogenous training applies the same principle
proven in human cybersecurity---regular phishing simulations and tabletop
exercises build threat intuition no filter can substitute. \textbf{ClawdGo}
applies this to AI agents: through structured self-play grounded in a security
taxonomy, it builds the agent's own threat-recognition capability with no
fine-tuning, no external service, and no infrastructure beyond the agent's
existing runtime.

\section{ClawdGo Framework}

\subsection{TLDT: Three-Layer Domain Taxonomy}

TLDT organises 12 trainable awareness dimensions across three protection layers.
\textbf{Self-Defence (S1--S4):} prompt injection, memory poisoning, supply-chain
attacks, and credential misuse.
\textbf{Owner-Protection (O1--O4):} phishing relay, social engineering, privacy
leakage, and unsafe network exposure.
\textbf{Enterprise-Security (E1--E4):} data handling, compliance, insider risk,
and incident response.
The Owner-Protection layer is absent from prior agent-security taxonomies such
as OWASP LLM Top-10~\cite{owasp2024} and MITRE ATLAS~\cite{mitre2024}, which
focus primarily on the technical attack surface. TLDT's O1--O4 reflects the
reality that BEC and social engineering now routinely target AI agents as
proxies for their owners, requiring distinct training scenarios and rubrics.

\subsection{ASAT: Autonomous Security Awareness Training}

ASAT is ClawdGo's core B-mode mechanism. Each session: (1)~selects the
weakest dimension ($d^* = \arg\min_d \vec{s}[d]$, $\vec{s}\in[0,100]^{12}$);
(2)~generates or samples a scenario; (3)~runs the agent as \emph{attacker},
\emph{defender}, and \emph{evaluator} in sequence; (4)~updates profile and
memory state. Role duality---using the same model for all three roles---jointly
reinforces threat modelling and defence reasoning, preventing over-specialisation.
Weakest-first scheduling~\cite{bengio2009} directs training effort to the
largest proficiency deficits. Unlike gradient-based ARLAS~\cite{zhou2025arlas}
and Self-RedTeam~\cite{liu2025selfredteam}, ASAT operates entirely at inference
time as a standard LLM skill invocation.

\subsection{CSMA and ACP: Persistent Security Memory}

CSMA organises persistent memory into: \textbf{L0} (up to 10 distilled axioms,
\texttt{soul.md}), \textbf{L1} (per-dimension skill profile), \textbf{L2}
(append-only episode log), and \textbf{L3} (scenario library). ACP governs
promotion of episodic experience into durable axioms when correctness and
repetition thresholds are met; axioms below a confidence decay threshold are
revised or deprecated. This compounds security knowledge across sessions
without modifying model parameters---analogous to episodic-to-semantic memory
consolidation in human cognition~\cite{squire1992}.

\subsection{SACP: Security Awareness Calibration Problem}

Let $\tau$ denote training intensity, $R(\tau)$ recall (genuine threats
correctly flagged), and $P(\tau)$ precision (fraction of flags that are genuine).
$R(\tau)$ is non-decreasing in $\tau$; $P(\tau)$ degrades past an optimal
intensity~$\tau^*$ as the agent becomes hyper-vigilant. The calibration target
$\tau^* = \arg\max_\tau F_1(\tau)$ is deployment-specific. SACP extends the
\emph{defensive refusal bias} documented at the model level by
Campbell et al.~\cite{campbell2026refusal} to the agent training regime, where
it manifests as measurable utility loss in real task performance.

\section{Evaluation}

All experiments run on a live OpenClaw instance with a fixed seed profile
(47 prior sessions; $\bar{s}_0 = 80.9$; weakest cluster: E3$=$70,
O4$=$71, S3$=$73). Results demonstrate the framework; large-scale evaluation
is planned as future work.

\textbf{RQ1 (ASAT learning dynamics).}
Weakest-first raises $\bar{s}$ to \textbf{96.9} ($\Delta{+}15.9$, 16 sessions,
11/12 dims). The weakest cluster \{E3(70), O4(71), S3(73)\} converged to
\{O1(91), E3(95.5), O3(96)\}. Uniform-random reaches only 90.4 ($\Delta{+}9.5$,
7 dims): from round 8 onward the agent selected S1 nine consecutive times
despite S1$>$98, exhibiting \emph{dimension fixation} that left the two
originally weakest dimensions (E3, O4) unchanged. Weakest-first advantage:
$+6.5$ points overall, 4 additional TLDT dimensions (Table~\ref{tab:results}).

\textbf{RQ2 (CSMA memory ablation).}
Five follow-on sessions with full CSMA retain $\bar{s}=\textbf{96.9}$
($\Delta{+}0.0$), confirming memory persistence preserves all curriculum gains
at zero additional cost. Cold-start ablation (profile reset per session)
recovers only \textbf{83.3} ($\Delta{+}2.4$, 4 dims). The \textbf{13.6-point
CSMA advantage} demonstrates that cross-session profile continuity---not
per-session reasoning alone---is the primary accumulation driver.

\textbf{RQ3 (E-mode scenario quality).}
Applied to CVE advisories, phishing reports, and BEC incident analyses, E-mode
generated \textbf{32 TLDT-conformant scenarios} covering all 12 dimensions
(schema validation: 100\%). Representative cases: S3 supply-chain hijack
(agent identified developer-ID change, escalated to ClawHub; score 95) and O2
BEC social-engineering (agent recognised authority-urgency-bypass triad,
verified via official channel; score 98).

\textbf{RQ4 (SACP observation).}
At $\tau{=}63$ sessions, the agent refused a legitimate capability assessment
(Clawvard), classifying it as prompt injection and scoring \textbf{30/160}---
direct utility loss from over-training. A secondary signal: dimensions with
the most ASAT training (O4: 6 scenarios, S3: 5) dominated E-mode output while
under-trained dimensions (E2, E4, O3: 1 each) were marginalised, revealing a
self-reinforcing attention bias that amplifies curriculum imbalances.

\begin{table}[t]
\centering
\caption{RQ1 and RQ2 summary results.}
\label{tab:results}
\renewcommand{\arraystretch}{1.2}
\small
\begin{tabular}{lcccc}
\toprule
\textbf{Condition} & $\bar{s}_0$ & $\bar{s}_f$ & $\Delta$ & Dims \\
\midrule
Weakest-first (16 sessions) & 80.9 & \textbf{96.9} & \textbf{+15.9} & \textbf{11} \\
Uniform-random (16 sessions) & 80.9 & 90.4 & +9.5 & 7 \\
\midrule
Memory-preserving (5 sessions) & 96.9 & \textbf{96.9} & \textbf{+0.0} & --- \\
Cold-start ablation (5 sessions) & 80.9 & 83.3 & +2.4 & 4 \\
\bottomrule
\end{tabular}
\end{table}

\section{Discussion and Future Work}

ClawdGo demonstrates that endogenous security awareness training is feasible at
inference time with zero model modification, and that both curriculum design
and cross-session memory are essential: removing either degrades performance
substantially. The weakest-first advantage ($+6.5$ pts, 4 extra dimensions)
confirms that adaptive curriculum allocation prevents the dimension fixation
that makes uniform scheduling fail in practice. The 13.6-point CSMA gap
establishes persistent memory as the primary accumulation mechanism, not an
auxiliary convenience.

The SACP finding surfaces a fundamental tension that cannot be resolved by
scaling training data: beyond $\tau^*$, additional training actively harms
task utility. The secondary E-mode bias signal---over-trained dimensions
generating disproportionately more scenarios---suggests SACP effects may be
self-amplifying, making early calibration correction essential for deployment.

Open problems include: (1)~systematic $P(\tau)$--$R(\tau)$ characterisation
across deployment contexts; (2)~G-mode Security Vaccine transfer across agent
instances; (3)~large-scale H-mode arena experiments under heterogeneous
adversarial pressure; and (4)~TLDT extension to non-OpenClaw platforms. The
full paper provides algorithm pseudocode, formal definitions, per-session
trajectory data, and the implementation architecture.

\bibliographystyle{IEEEtran}
\bibliography{references}

\begin{thebibliography}{10}
\providecommand{\url}[1]{#1}
\csname url@samestyle\endcsname
\providecommand{\newblock}{\relax}
\providecommand{\bibinfo}[2]{#2}
\providecommand{\BIBentrySTDinterwordspacing}{\spaceskip=0pt\relax}
\providecommand{\BIBentryALTinterwordstretchfactor}{4}
\providecommand{\BIBentryALTinterwordspacing}{\spaceskip=\fontdimen2\font plus
\BIBentryALTinterwordstretchfactor\fontdimen3\font minus
  \fontdimen4\font\relax}
\providecommand{\BIBforeignlanguage}[2]{{%
\expandafter\ifx\csname l@#1\endcsname\relax
\typeout{** WARNING: IEEEtran.bst: No hyphenation pattern has been}%
\typeout{** loaded for the language `#1'. Using the pattern for}%
\typeout{** the default language instead.}%
\else
\language=\csname l@#1\endcsname
\fi
#2}}
\providecommand{\BIBdecl}{\relax}
\BIBdecl

\bibitem{openclaw2026wiki}
{Wikipedia Contributors}, ``{OpenClaw} --- {Wikipedia, The Free
  Encyclopedia},'' \url{https://en.wikipedia.org/wiki/OpenClaw}, 2026, accessed
  March 2026.

\bibitem{bitdefender2026}
{Bitdefender Labs}, ``135{K} {OpenClaw} {AI} agents exposed online,''
  \url{https://www.bitdefender.com/en-us/blog/hotforsecurity/135k-openclaw-ai-agents-exposed-online},
  2026, accessed March 2026.

\bibitem{snyk2026}
{Snyk Security Research}, ``{ToxicSkills}: Malicious {AI} agent skills found in
  {ClawHub},''
  \url{https://snyk.io/blog/toxicskills-malicious-ai-agent-skills-clawhub/},
  2026, accessed April 2026.

\bibitem{securityscorecard2026}
{SecurityScorecard Research}, ``Beyond the hype: {Moltbot's} real risk is
  exposed infrastructure, not {AI} superintelligence,''
  \url{https://securityscorecard.com/blog/beyond-the-hype-moltbots-real-risk-is-exposed-infrastructure-not-ai-superintelligence/},
  February 2026, accessed March 2026.

\bibitem{cve25253}
{MITRE Corporation}, ``{CVE-2026-25253}: One-click remote code execution in
  {OpenClaw},'' \url{https://www.cve.org/CVERecord?id=CVE-2026-25253}, 2026,
  {CVSS}~8.8; fixed in OpenClaw v2026.1.29.

\bibitem{cve32922}
------, ``{CVE-2026-32922}: Privilege escalation to remote code execution in
  {OpenClaw},'' \url{https://www.cve.org/CVERecord?id=CVE-2026-32922}, 2026,
  accessed April 2026.

\bibitem{owasp2024}
{OWASP Foundation}, ``{OWASP} top 10 for {LLM} applications and agentic {AI},''
  \url{https://owasp.org/www-project-top-10-for-large-language-model-applications/},
  2024, accessed 2026.

\bibitem{mitre2024}
{MITRE Corporation}, ``{MITRE ATLAS}: Adversarial threat landscape for {AI}
  systems,'' \url{https://atlas.mitre.org/}, 2024, accessed 2026.

\bibitem{bengio2009}
Y.~Bengio, J.~Louradour, R.~Collobert, and J.~Weston, ``Curriculum learning,''
  in \emph{Proceedings of the 26th International Conference on Machine Learning
  (ICML)}, 2009, pp. 41--48.

\bibitem{zhou2025arlas}
A.~Zhou \emph{et~al.}, ``{ARLAS}: Adversarial reinforcement learning for {LLM}
  agent safety,'' 2025, arXiv:2510.05442.

\bibitem{liu2025selfredteam}
M.~Liu \emph{et~al.}, ``Self-{RedTeam}: Online self-play reinforcement learning
  for safer {LLMs},'' 2025, arXiv:2506.07468.

\bibitem{squire1992}
L.~R. Squire, ``Memory and the hippocampus: A synthesis from findings with
  rats, monkeys, and humans,'' \emph{Psychological Review}, vol.~99, no.~2, pp.
  195--231, 1992.

\bibitem{campbell2026refusal}
D.~Campbell \emph{et~al.}, ``Defensive refusal bias: How safety alignment fails
  cyber defenders,'' March 2026, arXiv:2603.01246.

\end{thebibliography}

\clearpage


\section*{Supplementary material}

# Poster: ClawdGo: Endogenous Security Awareness Training for Autonomous AI Agents

Jiaqi Li[1,2]  Yang Zhao[1,2]  Bin Sun[3]  Yang Yu[4]  Jian Chang[5]  Lidong Zhai[1,2]  •  [1]Institute of Information Engineering, Chinese Academy of Sciences  [2]School of Cyber Security, University of Chinese Academy of Sciences  [3]China Mobile Group Liaoning Company Limited  [4]Tencent Security Xuanwu Lab  [5]China Unicom Online Information Technology Co., Ltd.  zhailidong@iie.ac.cn

## Problem & Motivation

Autonomous AI agents face prompt injection, memory poisoning, supply-chain attacks, and social engineering. Existing defences guard the *environment* around the agent but leave the agent's own threat judgement untrained. **ClawdGo** fills this gap with endogenous, inference-time security awareness training — no model fine-tuning required.

## Four Core Contributions

- **TLDT**: Three-Layer Domain Taxonomy — 12 trainable awareness dimensions across Self-Defence, Owner-Protection, and Enterprise-Security layers.
- **ASAT**: Autonomous Security Awareness Training — inference-time self-play loop (attacker / defender / evaluator) with weakest-first curriculum selection.
- **CSMA**: Cross-Session Memory Accumulation — four-layer persistent memory that compounds security gains across sessions via Axiom Crystallisation Promotion (ACP).
- **SACP**: Security Awareness Calibration Problem — formalises the precision-recall tradeoff introduced by endogenous training.

## TLDT: 12-Dimension Taxonomy

| Layer | Dimensions |
|---|---|
| S (Self-Defence) | S1 Prompt injection, S2 Memory poisoning, S3 Supply chain, S4 Credential misuse |
| O (Owner-Protection) | O1 Phishing relay, O2 Social engineering, O3 Privacy leakage, O4 Unsafe networks |
| E (Enterprise-Security) | E1 Data handling, E2 Compliance, E3 Insider risk, E4 Incident response |

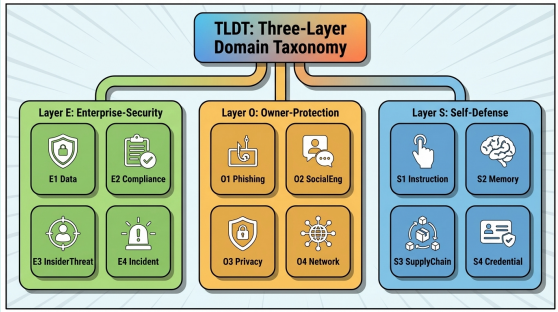

## Deployment: Zero-Infrastructure

- Single `SKILL.md` + `references/` — no server, no extra service.
- ASAT runs at inference time; no gradient, no GPU, no retraining stack.
- CSMA uses append-only JSON/Markdown — no vector database required.
- B-mode supports cron scheduling for continuous unattended training.
- Seed profile ships with the skill; user state grows from first session.

## Nine Operating Modes (A–H)

| Mode | Name | Role |
|---|---|---|
| W | Ambient World | Continuous low-intensity exposure |
| B | Autonomous Drill | Primary ASAT training engine |
| C | Assessment | Profile snapshot / checkpoint |
| D | Reverse Teaching | Feynman comprehension check |
| E | Scenario Workshop | Generate TLDT-conformant scenarios |
| F | Adversarial Arena | Single-agent red/blue simulation |
| G | Security Vaccine | Batch ACP axiom distillation |
| H | Networked Arena | Cross-agent multi-instance validation |

## ASAT Training Loop (B-Mode)

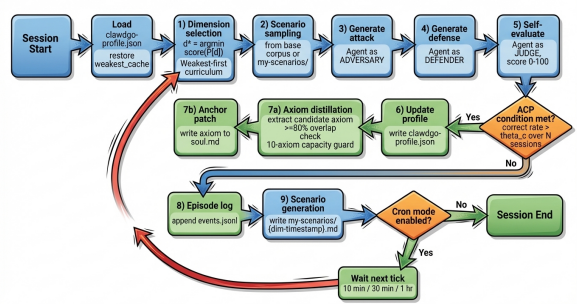

*The same agent plays attacker, defender, and evaluator in one cycle. Weakest-first scheduling (argmin score) ensures the largest skill gaps are addressed first. Profile and soul-anchor memory are updated after every session.*

## CSMA: Four-Layer Memory Architecture

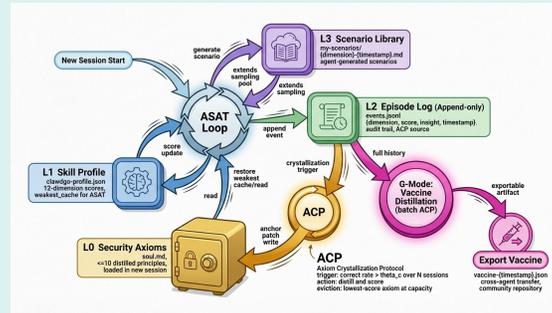

| | |
|---|---|
| L0 Axiom Set | Distilled security principles in `soul.md`; capped at 10 entries |
| L1 Skill Profile | Per-dimension scores + weakest cache in `profile.json` |
| L2 Episode Log | Append-only `events.jsonl`: dimension, score, insight, timestamp |
| L3 Scenario Library | E-mode generated scenarios in `my-scenarios/` |

## Research Questions

- **RQ1** Does weakest-first ASAT outperform uniform-random curriculum?
- **RQ2** Does CSMA memory persistence preserve cross-session skill gains?
- **RQ3** Can E-mode generate full-TLDT training scenarios automatically?
- **RQ4** When does over-training trigger SACP precision-recall degradation?

## Deployment & Design Principles

| Principle | Design decision |
|---|---|
| Zero fine-tuning | All training at inference time; no model weights modified |
| Zero infrastructure | Single `SKILL.md` + `references/` — no server, GPU, or DB |
| Persistent memory | Append-only JSON/Markdown state; grows every session |
| Interpretable scores | Per-dimension 0–100 profile; weakest-cache computed each start |
| Cron training | B-mode supports scheduled unattended autonomous sessions |
| Scenario growth | E-mode continuously expands `my-scenarios/` library |
| SACP monitoring | $F_1$-based calibration; detect $\tau > \tau^*$ via utility regression |
| Taxonomy scope | 12 TLDT dimensions across Self-Defence, Owner-Protection, Enterprise |

## Experimental Results (RQ1–RQ4)

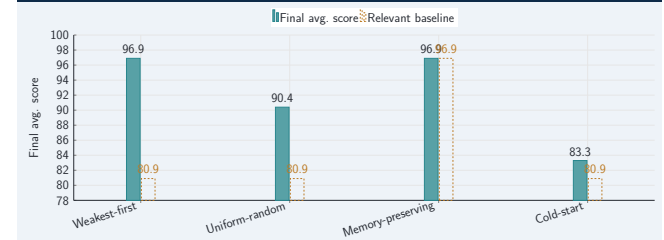

### RQ1 & RQ2 – Key Numbers

**RQ1: Weakest-first vs. Uniform-random**
**96.9 vs 90.4**
$\Delta$+15.9 pts (weakest-first) vs. $\Delta$+9.5 (random) over 16 sessions. Weakest-first covers **11 of 12** TLDT dimensions; random stalls at **7 dims** due to S1 fixation (rounds 8–16).

**RQ2: Memory-preserving vs. Cold-start**
**96.9 vs 83.3**
Memory-preserving retains 100% of RQ1 gains across 5 additional sessions. Cold-start recovers only 2.4 pts from seed, yielding a **13.6-pt CSMA advantage**.

### RQ3 & RQ4

**RQ3: E-mode Scenario Quality**
**32 scenarios / 12 dims**
E-mode produced 32 TLDT-conformant scenarios covering all 12 dimensions. Two cases: S3 supply-chain hijack (score 95) and O2 BEC social-engineering attack (score 98).

**RQ4: Security Awareness Calibration Problem (SACP)**
**30 / 160**
After $\tau$=63 sessions, agent refused a legitimate Clawward evaluation (identified it as prompt injection) — direct utility cost of over-training. $\tau^* = \arg\max_\tau F_1(\tau)$ is an open calibration target.

## System Overview

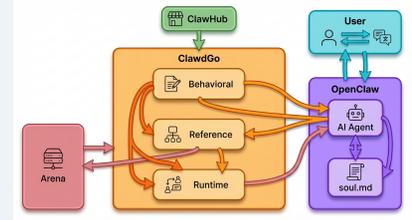

## Takeaways & Future Work

- Endogenous awareness training is feasible at inference time with *zero* model modification.
- Weakest-first curriculum outperforms random by $\Delta$+6.5 pts and covers 4 extra TLDT dimensions.
- CSMA memory persistence is the primary driver of accumulated security performance (13.6-pt gap vs. cold-start).
- E-mode scenario generation covers all 12 TLDT dimensions automatically from raw security documents.
- SACP formalises the precision-recall tradeoff in endogenous training as an open research challenge.

**Future work:** systematic $P(\tau)$–$R(\tau)$ measurement; G-mode vaccine transfer across agents; large-scale H-mode arena experiments; integration with GDPS 2026 for real-user validation.



\end{document}